\documentclass[aps,twocolumn,prd,nofootinbib]{revtex4}

\usepackage{amsmath}
\usepackage{graphicx}
\usepackage{dcolumn}
\usepackage{bm}
\usepackage{amssymb}
\usepackage{latexsym}

\def\be{\begin{equation}}
\def\ee{\end{equation}}
\def\ba{\begin{eqnarray}}
\def\ea{\end{eqnarray}}

\begin{document}

\title{Phantom Inflation in Little Rip}

\author{Zhi-Guo Liu\footnote{Email: liuzhiguo08@mails.gucas.ac.cn}}
\author{Yun-Song Piao\footnote{Email: yspiao@gucas.ac.cn}}

\affiliation{College of Physical Sciences, Graduate School of
Chinese Academy of Sciences, Beijing 100049, China}

\begin{abstract}


We study the phantom inflation in little rip cosmology, in which the
current acceleration is driven by the field with the parameter of
state $w < -1$, but since $w$ tends to -1 asymptotically, the rip
singularity occurs only at infinite time. In this scenario, before
the rip singularity is arrived, the universe is in an inflationary
regime. We numerically calculate the spectrum of primordial
perturbation generated during this period and find that the results
may be consistent with observations. This implies that if the
reheating happens again, the current acceleration might be just a
start of phantom inflation responsible for the upcoming
observational universe.

\end{abstract}

\maketitle


\section{Introduction}


The phantom field, in which the parameter of the equation of state
$w<-1$, has still acquired increasingly attention \cite{RC},
inspired by the study of dark energy, e.g.
\cite{CKW},\cite{SSD},\cite{HL},\cite{ENO},\cite{GPZ},\cite{Chimento},\cite{KSS},\cite{VF},\cite{Bamba}.
The actions with phantomlike behavior may be arise in supergravity
\cite{N}, scalar tensor gravity \cite{BEPS}, higher derivative
gravity \cite{P}, braneworld
 \cite{SS}, string theory \cite{FM}, and
other scenarios \cite{CL,S}, and also from quantum effects
\cite{Onemli}. The universe driven by the phantom field will
evolve to a singularity, in which the energy density become
infinite at finite time, which is called the big rip, see
Refs.\cite{NOT},\cite{Barrow},\cite{S1},\cite{BG},\cite{Dabrowski},\cite{BGM}
for other future singularities .

Recently, the little rip scenario has been proposed \cite{FLS}, in
which the current acceleration of universe is driven by the
phantom field, the energy density increases without bound, but $w$
tends to -1 asymptotically and rapidly, and thus the rip
singularity dose not occur within finite time. This scenario still
leads to a dissolution of bound structures at some epochs
\cite{FLS,FLS2}, which is similar to the big rip, however, in
little rip scenario, the universe arrives at the singularity only
at infinite time. This has motivated some interesting studies
\cite{BEN}, and also asymptotically phantom dS universes or
pseudo-rip \cite{Prip},\cite{Atv}.

The simplest realization of phantom field is the scalar field with
reverse sign in its dynamical term. In little rip scenario,
initially $w<-1$, the energy density of the phantom field will
increase with time, and arrive at a high energy scale at finite
time, and at this epoch the phantom field has $w\simeq -1$. Thus
after some times or before the rip singularity of universe is
arrived, if the exit from the little rip phase and the reheating
happen, the inflation, which is driven by the phantom field, in
little rip phase might be responsible for the upcoming
observational universe.

The phantom inflation has been proposed in \cite{YZ}, and also in
\cite{phan},\cite{GJ},\cite{BI},\cite{NO},\cite{BFM},\cite{WY},\cite{Y},\cite{Fe},\cite{Liu1012}.
Recently, there has been a relevant study in modified gravity
\cite{Toprensky}. In phantom inflation scenario, the power spectrum
of curvature perturbation can be nearly scale invariant. The duality
of primordial spectrum to that of normal field inflation has been
studied in \cite{phan, Piao2004}. The exit from the phantom
inflation has been studied
\cite{YZ},\cite{GJ},\cite{WY},\cite{Y},\cite{Fe}. The spectrum of
tensor perturbation in phantom inflation is slightly blue tilt
\cite{YZ},\cite{Piao0601}, which is distinguished from that of
normal inflationary models.


Here, we will argue that after a finite period of the little rip
phase the universe might recur the evolution of observational
universe. We will calculate the spectrum of primordial
perturbation during the phantom inflation in various little rip
phases, and find that the spectrum may be consistent with
observations. 
We discussed the implication of this result in the final.



\section{The phantom inflation in little rip}

In little rip phase, initially $w<-1$, the energy density of the
phantom field will increase with time, and arrive at a high energy
scale at finite time, and at this epoch the phantom field has
$w\simeq -1$. Here, we actually required that before $w\simeq -1$,
the phantom field must have arrived at a high energy regime, which
assures the occurrence of inflation. In general, this can be
implemented by introducing a monotonically increasing potential of
the phantom field.

The phantom field satisfies the equations
\ba 3H^2 & =& -\frac{1}{2}\dot{\phi}^2+V(\phi),\label{H}\\
\ddot{\phi} & + & 3H\dot{\phi}-V_{\phi}(\phi)=0, \label{phi} \ea
where $M_P^2/8\pi=1$, and the dot is the derivative with respect
to the physical time $t$. $H^2>0$ requires that in all case for
the phantom evolution $\dot{\phi}^2$ must be smaller than its
potential energy.
The phantom field will be driven to climb up along its potential,
which is reflected in the minus before $V_{\phi}$ term. We define
the slow climbing parameters \cite{YZ}, as for normal inflation.
\begin{eqnarray}
\epsilon_{Pha}\equiv -\frac{\dot H}{H^2}<0~~~,~~~
\delta_{Pha}\equiv-\frac{\ddot\phi}{H\dot\phi}.
\end{eqnarray}
Thus the equation of state is \be w_{Pha} ={2\over
3}\epsilon_{Pha}-1. \ee

\begin{figure}[htbp]
\includegraphics[scale=0.6,width=8.5cm]{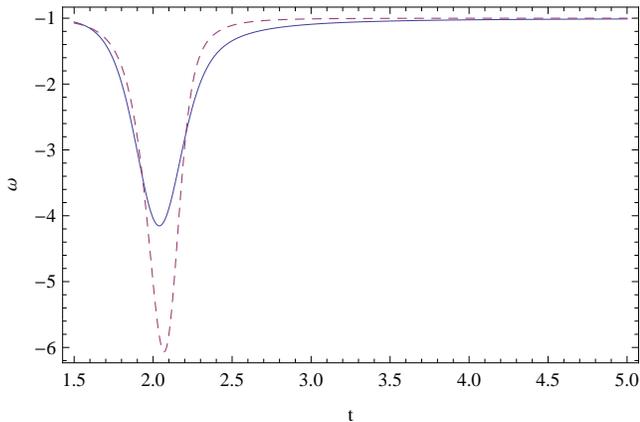}
\caption{The evolutions of $w$ with the time for $V(\phi)\sim
\phi^2$ (solid line) and $\phi^3$ (dashed line), respectively, where
$\phi$ is the phantom field, which is initially around the bottom of
its potential. } \label{fig:h}
\end{figure}

We consider the monotonically increasing potential of the phantom
field as $V(\phi)\sim \phi^N$. The phantom field initially drives
the current acceleration, and is in low energy regime with
$w\simeq -1$. Then the field will climb up along its potential,
$w$ will rapidly deviate from $-1$ during this period. However,
when the phantom field climbs to $\phi\simeq 1$,
$|\epsilon_{Pha}|\sim 1$. Hereafter, the conditions
$|\epsilon_{Pha}|<1$ and $|\delta_{Pha}|<1$ are satisfied, since
\be \left|\epsilon_{Pha}\right|=\frac{\dot H}{H^2}\sim {N^2\over
\phi^2}, \ee which is smaller for larger $\phi$. Thus $w\simeq -1$
is obtained again, which is the phantom inflationary phase, since
the phantom field is in high energy regime. Thus in the little rip
cosmology, it is generally that before the the universe arrives at
the singularity, it will enter into a period of the phantom
inflation, up to the singularity at infinite time.

\begin{figure}[htbp]
\includegraphics[scale=0.6,width=8.0cm]{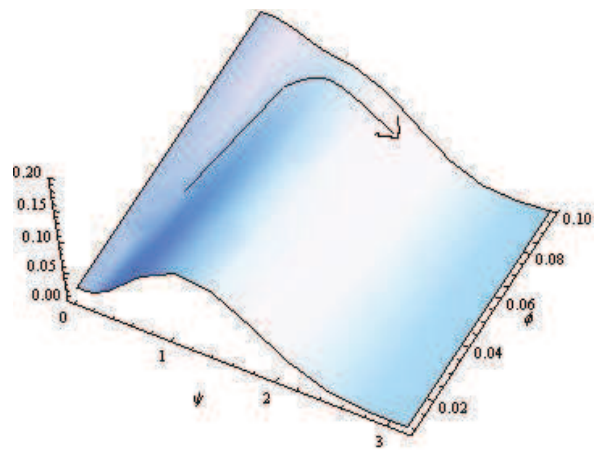}
\caption{The sketch of the exit from little rip phase. $\phi$ is the
phantom field, while $\psi$ is the normal waterfall field, as that
in hybrid inflation \cite{linde94},\cite{hybrid1}. Initially $\phi$
is at the bottom of its potential, which then climbs up along its
potential. The effective $M^2_{\psi}$ of $\psi$ is initially
positive, which will gradually decrease with the increasing of
$\phi$. When $\psi$ becomes tachyonical, and $\psi$ will rapidly
roll down along its potential. Thus at this time almost all energy
of $\phi$ will be released.  } \label{fig:exit}
\end{figure}

The evolutions of $w$ are plotted in Fig.\ref{fig:h} for the
potential $\phi^N$. The result is consistent with above
discussions.


However, it is possible that some times after the energy density
of the phantom field arrives at a high energy scale, or before the
rip singularity of universe is arrived, the energy of field will
be released, and the universe reheats, after which the evolution
of hot ``big bang" recurs. The exit mechanism in Ref.\cite{YZ} is
depicted in Fig.\ref{fig:exit}.

However, before the phantom field arrives at the high energy regime
responsible for inflation, if the exit from the little rip phase or
the reheating happens, the phantom inflation will not exist. Thus
whether the phantom inflation occur will depend on the exiting scale
from little rip phase. Here, we require this exiting scale is at
least about $M\sim \rho_{inf}^{1/4}\sim 10^{-3}$, $\rho_{inf}\sim
10^{-12}$ is the inflationary energy density. Before this, the
current bound structures have disintegrated, see Appendix A.

\section{The spectrum of primordial perturbation }

\subsection{Basic results}

We will calculate the spectrum of curvature perturbation during
the phantom inflation. We define \cite{M},\cite{KS} \be u = -z
\mathcal{R}\ee where $\cal R$ is the curvature perturbation, and
$z\equiv a\sqrt{2|\epsilon_{Pha}|}$.

The motive equation of $u_{\rm k}$ in the $k$ space is
\begin{equation}  \label{uk}
u_{\rm k}'' + \left( k^2-\frac{z''}{z} \right)u_{\rm k}=0,
\end{equation}
where the prime is the derivative with respect to conformal time.
In general, when $k^2 \gg z''/z$, the solution of perturbation is
\begin{equation}  \label{freesc}
u_{{\rm k}} \sim\frac{1}{\sqrt{2k}} \, e^{-ik\eta}\,.
\end{equation}
The $\frac{1}{\sqrt{2k}}$ is obtained by the normal quantization
of mode function $u_k$, which seems illogical for phantom field,
since there is a ghost instability induced by $\epsilon<0$.
However, it might be thought that the evolution with $w<-1$
emerges only for a period, which may be simulated
phenomenologically by the phantom field, while the full theory
should be well behaved. It has been showed in Ref.\cite{Piao1105}
that $\epsilon$ can be replaced with $|\epsilon|$, due to the
introduction of generalized Galileon field
\cite{Vikman},\cite{KYY}, there is not the ghost instability, the
perturbation theory is healthy.

While $k^2 \ll z''/z$, the dominated mode is
\begin{equation}
u_{{\rm k}} \sim z\ \label{eqn:regmode}
\end{equation}
which means that the curvature perturbation
$|{\mathcal R}_{{\rm k}}|=|u_{{\rm k}}/z|$
is constant in this regime.

The power spectrum ${\cal P}_{{\cal R}}(k)$ is defined as
\begin{equation}
\langle {\cal R}_{{\rm k}_1} {\cal R}^*_{{\rm k}_2} \rangle =
        \frac{2\pi^2}{k^3} {\cal P}_{{\cal R}} \delta^{3} \,
        (k_1- k_2) \,,
\end{equation}
and is given by
\begin{equation}
\label{pspec} {\cal P}_{{\cal R}}^{1/2}(k) =
\sqrt{\frac{k^3}{2\pi^2}} \,
        \left| \frac{u_{{\rm k}}}{z} \right| \,.
\end{equation}

In slow climbing approximations $|\epsilon_{Pha}|\ll 1$ and
$|\delta_{Pha}|\ll 1$,
\begin{eqnarray} \mathcal {P}_\mathcal
{R}^{1/2}=\frac{1}{\sqrt{2|\epsilon_{Pha}|}}\frac{H}{2\pi},
\label{p}\end{eqnarray} where the spectral index is \be n_\mathcal
{R}-1\simeq -4\epsilon_{Pha}+2\delta_{Pha}.\ee Noting the spectral
index is only determined by $\epsilon_{Pha}$ but not its absolute
value, since we have
$|\epsilon_{Pha}|^\prime/|\epsilon_{Pha}|=\epsilon_{Pha}^\prime/\epsilon_{Pha}$
and
$|\epsilon_{Pha}|^{\prime\prime}/|\epsilon_{Pha}|=\epsilon_{Pha}^{\prime\prime}/\epsilon_{Pha}$
in the calculation of $\frac{z''}{z}$.

This spectrum may be either blue or red. The results are dependent
on the relative magnitude of $\epsilon_{Pha}$ and $\delta_{Pha}$,
see Ref.\cite{YZ} for the details.

In big rip scenario, the big rip singularity requires
$w_{Pha}<-1$, the primordial spectrum obtained is not scale
invariant, unless $w_{Pha}$ is not far from $-1$. While in little
rip cosmology, $w_{Pha}$ tends to -1, thus with the increasing of
time we certainly have $|\epsilon_{Pha}|\ll 1$, which naturally
insure the scale invariance of spectrum.

\subsection{Numerical results}

Here, we will calculate ${\cal P}_{{\cal R}}$ by evolving the full
mode equation (\ref{uk}) numerically without any approximations.

We define $\alpha=\ln{a}$, and with this replacement, we have

\begin{eqnarray}\label{HH}
H_\alpha=\frac{1}{2}H\phi_\alpha^2,
\end{eqnarray}

\begin{eqnarray}\label{phi}
\phi_{\alpha\alpha}+(\frac{H_\alpha}{H}+3)\phi_\alpha-\frac{1}{H^2}V_{\phi}=0.
\end{eqnarray}
where the subscript $\alpha$ is the derivative for $\alpha$, which
will facilitate the numerical integration. The perturbation equation
(\ref{uk}) can be written as
\begin{eqnarray}
u_{\alpha\alpha}+(1+\frac{H_\alpha}{H})u_\alpha+\left(\frac{k^2}{e^{2\alpha}H^2}-\frac{z''/z}{e^{2\alpha}H^2}\right)u=0,
\label{ualpha}
\end{eqnarray}
where $\frac{z''}{z}$ is only the function of $H$ and its higher
order derivative with respect to $\alpha$, which here will be
determined by the numerical calculation.

It has been clearly explained in Refs.\cite{FLS},\cite{FLS2} that
before the rip singularity is arrived, the little rip universe
experiences the disintegration of bound structures. Hence,
actually in finite time, the universe becomes totally empty.
However, the asymptotical behavior of the evolution of $a$ is
significant to determine the character of perturbation spectrum
generated during this period.


The solution of Eq.(\ref{ualpha}) is only determined by the behavior
of $a$. Thus in principle, for any parametrization of $a$ depicting
the little rip phase, the corresponding power spectrum can be
calculated. The details of the implement of field theory dose not
significantly alter the result of Eq.(\ref{ualpha}), unless its
speed of sound is rapidly changed. Below, we will focus on some
interesting parametrizations of $a$. We will study some cases with
the phantom potential in Appendix B.

We firstly consider the parametrization of little rip in
Refs.\cite{FLS},\cite{FLS2} \be H=\,H_0 e^{\lambda t},
\label{Hp}\ee where $H_0$ and $\lambda$ are constant, which
implies \ba a & \sim & e^{{H_0\over \lambda}e^{\lambda t}},\\ w &
= & -1-{2\lambda\over 3H_0}e^{-\lambda t}.\ea Thus $w\simeq -1$
for $\lambda t\gg 1$, and there is not curvature singularity at
finite time. In Ref.\cite{FLS}, in low energy regime the
parameters chosen can make a best fit to the latest supernova
data. However, we are interested in the evolution in a high energy
regime with $w\simeq -1$, in which $\lambda t\gg 1$.

\begin{figure}[htbp]
\includegraphics[scale=0.6,width=8.0cm]{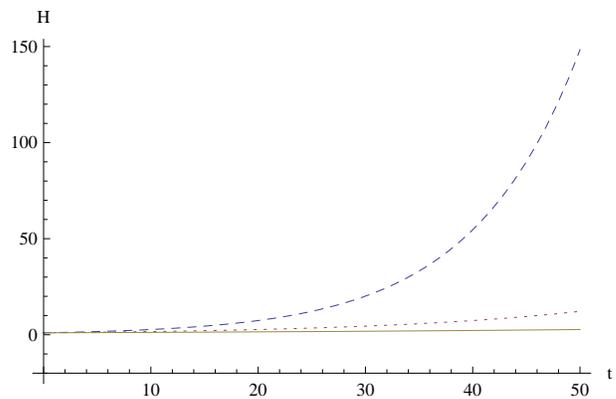}
\caption{The evolutions of $H$ in Eq.(\ref{Hp}) for $\lambda =0.1$
(the dashed line), 0.05 (the dotted line), 0.02 (the solid line),
respectively.} \label{fig:H1}
\end{figure}

\begin{figure}[htbp]
\includegraphics[scale=0.6,width=8.0cm]{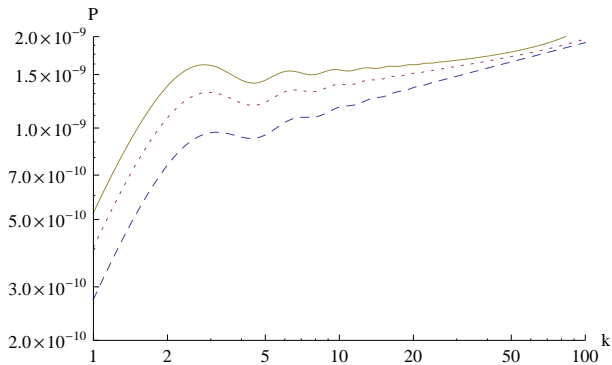}
\caption{The power spectrum of curvature perturbation corresponding
to the evolutions of $H$ in Fig.(\ref{fig:H1}), respectively. }
\label{fig:HH}
\end{figure}

The evolution of $H$ with this parametrization is plotted in
Fig.\ref{fig:H1}. The power spectrum is plotted in
Fig.\ref{fig:HH}. It seems that the spectrum is nearly scale
invariant only for small $\lambda$. However, this result is
dependent on the time interval selected. In principle, as long as
$\lambda t$ is enough large, the spectrum can be scale invariant
for any $\lambda$. It seems there is a cutoff $k_{cutoff}$ in
Fig.\ref{fig:HH}, however, which appears is because initially
there are some perturbation modes outside of the horizon. This
cutoff can be changed with the difference of the initial
parameters in the numerical calculation.

\begin{figure}[htbp]
\includegraphics[scale=0.6,width=6.0cm]{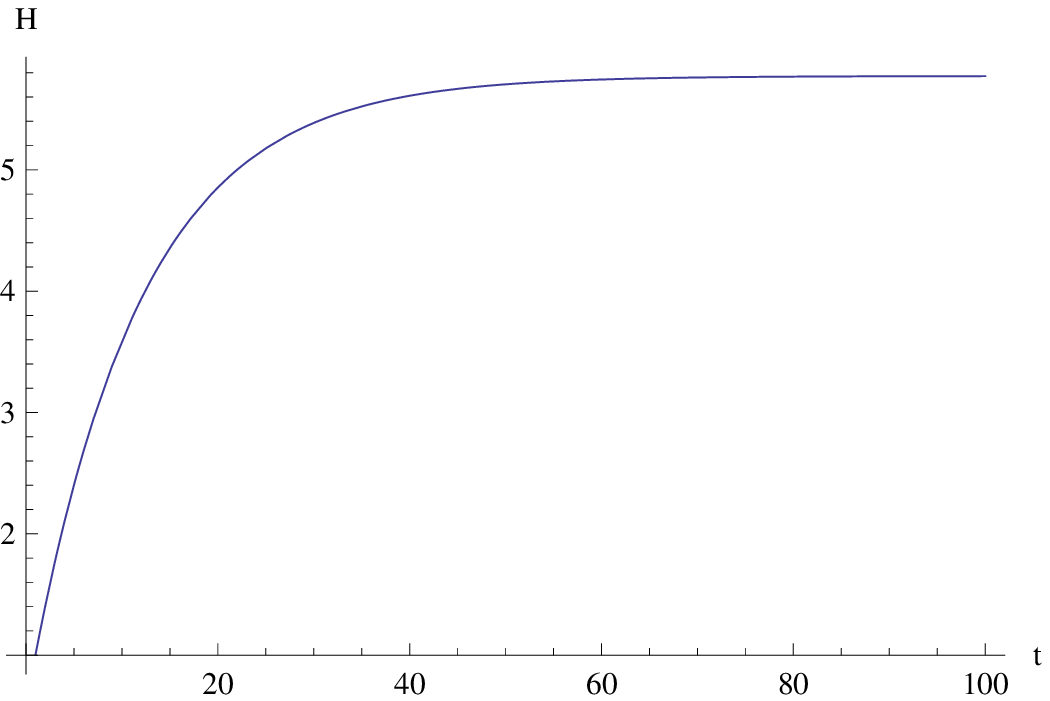}
\includegraphics[scale=0.6,width=6.0cm]{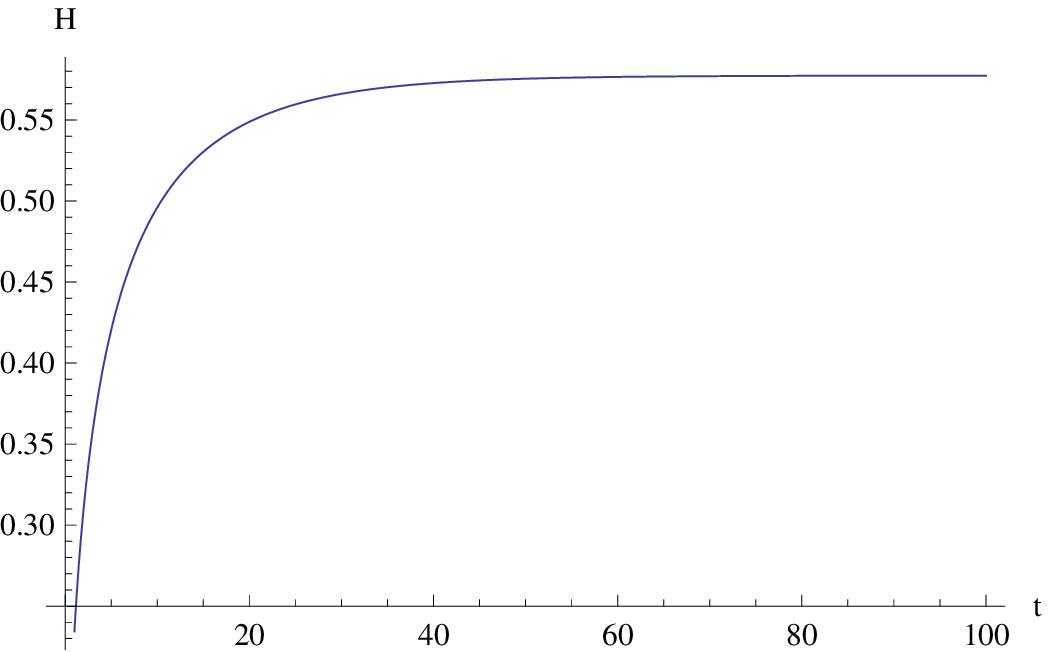}
\caption{The evolution of $H$. The upper panel corresponds to the
parametrization (\ref{a21}) for $x_0=1,b=1, x_f=10,$ and the lower
panel corresponds to the parametrization (\ref{a22}) for
$x_f=0.001,\gamma=10$ } \label{fig:H12}
\end{figure}

\begin{figure}[htbp]
\includegraphics[scale=0.6,width=8.0cm]{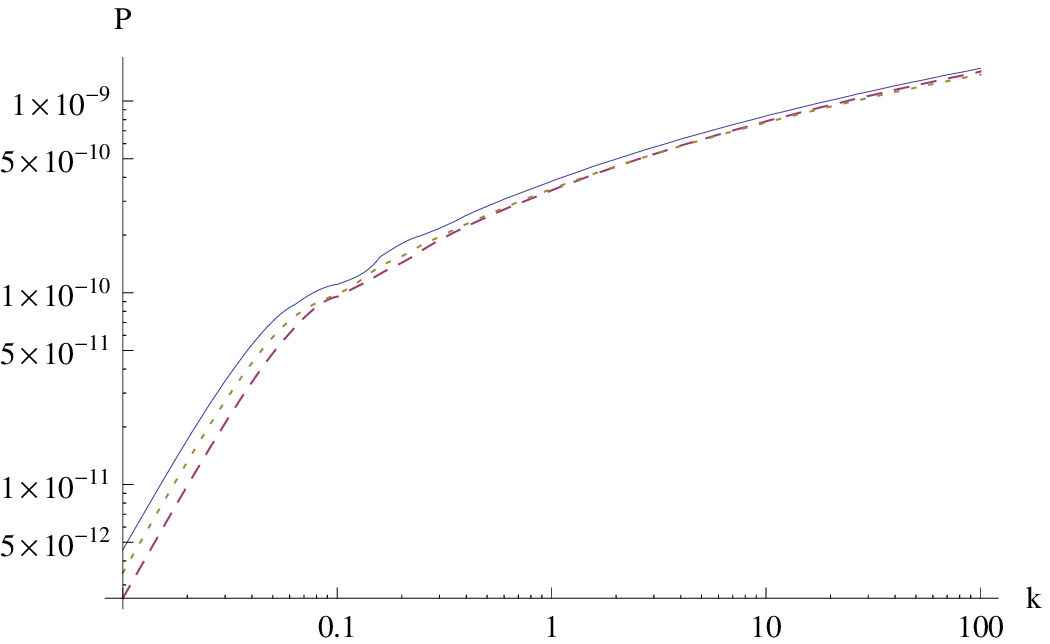}
\includegraphics[scale=0.6,width=8.0cm]{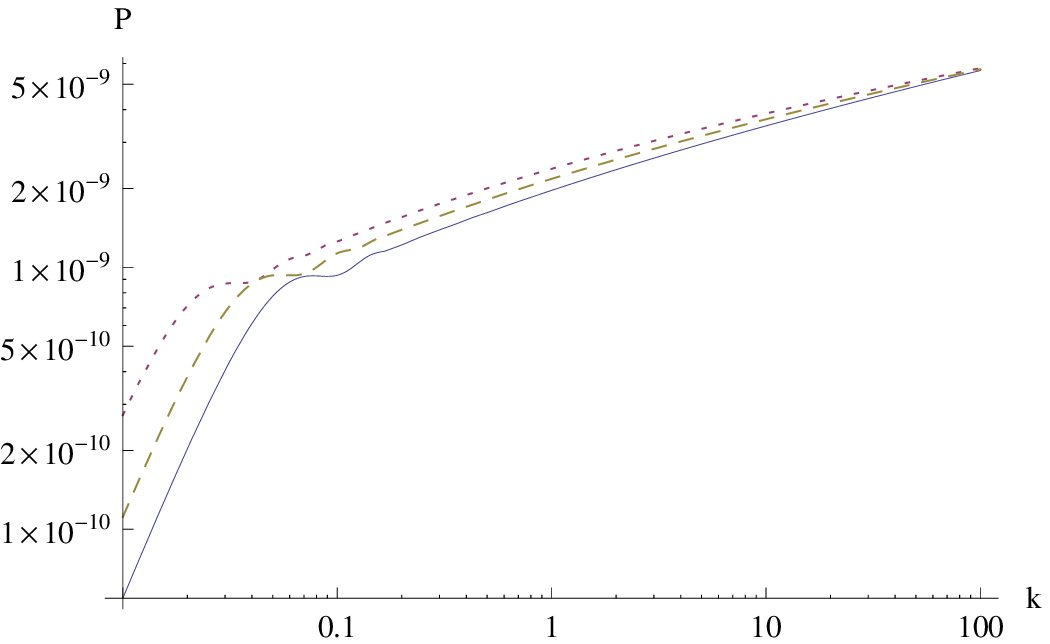}
\caption{The power spectrum of the curvature perturbation. The
upper panel correspond to the parametrization (\ref{a21}) for
$b=1, x_f=5$ (solid line), $b=1, x_f=10$(dotted line),$b=1.5,
x_f=4$(dashed line) and the lower panel correspond to the
parametrization (\ref{a22}) for $x_f=0.001$ and $\gamma=10$(solid
line), $\gamma=15$(dashed line),$\gamma=20$ (dotted line) ,
respectively.} \label{fig:pa12}
\end{figure}

Then we consider the parametrizations of $a$ in e.g.Ref.\cite{Atv}
for asymptotically phantom dS universes. The first is \ba a(t)\sim
e^{\frac{x_ft}{\sqrt{3}}}e^{g(t)},\label{a21}\ea \ba
g(t)=\frac{2x^2_f}{3b}(1-\frac{x_0}{x_f})(e^{-\frac{\sqrt{3}b
t}{2x_f}}-1), \ea where $x_0,x_f, b$ are constant and $x_f>x_0$.
The second is
\ba a(v)\sim \frac{1}{(1-v^{\frac{3}{2}})^{2/3\gamma}},
\label{a22}\ea
\ba t(v)=t_0-\frac{2}{\sqrt{3}\gamma x_f}ln(1-v),\ea
where $x_f, \gamma$ are constant, and the parameter $v$ varies
from $0 (t=t_0)$ to $1 (t\rightarrow \infty)$. In Ref.\cite{Atv},
the realizations of field theory have been given. Here, what we
require is only the evolutions of $a$. In fact, both cases have
similar behavior. The parameter $H$ grows at the beginning and
asymptotically tends to constant at late time, see
Fig.\ref{fig:H12}. The power spectrum is plotted in
Fig.\ref{fig:pa12}. The spectrums are nearly scale invariant and
are slightly blue tilt. We see that the smaller $x_f$ is for the
parametrization (\ref{a21}), the flatter the spectrum is, and the
smaller $\gamma$ is for the parametrization (\ref{a22}), the
flatter the spectrum is.

We see that for all considered parametrizations, the power spectra
of perturbations are nearly scale invariant, but slightly blue
tilt. The results are determined by the asymptotical behavior of
the evolution of $a$, which reflects how the state parameter $w$
the phantom field approaches $-1$.
In principle, there might be a classification of the
parametrization of $a$ in term of its giving the spectral tilts,
which might be interesting for further studies.

%

\section{Conclusion and Discussion}

In effective actions of some theories, the phantom field naturally
appear, which might be a suitable simulation of a fundamental
theory below certain physical cutoff. The cosmology with the
phantom field has been widely studied.

In little rip scenario, the current acceleration of universe is
driven by the phantom field with $w < -1$, the energy density of
the phantom field will increase without bound, but since $w$ tends
to -1 asymptotically, there is not the rip singularity at finite
time. This indicates that in little rip phase, the phantom field
will inevitably arrive at a high energy scale at late time, and at
this time it can have $w\simeq -1$, which corresponds to a period
of phantom inflation. Thus if the exit and the reheating happens
before the rip singularity of the universe is arrived, the phantom
inflation in little rip phase might be responsible for the coming
observational universe. Here, we have showed this possibility.

We have calculated the spectrum of primordial perturbation during
phantom inflation in different little rip scenario and
asymptotically phantom dS universes, and found that the results
may be consistent with observations. In normal inflationary
models, the spectrum of tensor perturbation is slightly red. While
the phantom inflation predicts a slightly blue spectrum of tensor
perturbation, which is distinguished from that of normal
inflationary models, but may be consistent with observations
\cite{Powell,ZH}.

Here, the calculating method for the primordial perturbation is
applicable for any parametrization of $a$. As long as $w\lesssim
-1$ is satisfied in high energy regime, the spectrum will be scale
invariant. However, in big rip scenario, the big rip singularity
requires $w_{Pha}<-1$, the primordial spectrum is scale invariant
only if $w_{Pha}$ is nearly around $-1$.

We can imagine that after the available energy of the phantom
field is released, it might be placed again in initial position of
its effective potential, and after the hot ``big bang" universe
evolves into the low energy regime, this field might dominate
again and roll up again along its potential, the universe comes
again to the little rip phase and the phantom inflation recurs.
This brings a scenario\cite{Feng06},\cite{Xiong08},\cite{IBF}, in
which the universe expands all along but the Hubble parameter
oscillates periodically. In this scenario, we live only in one
cycle, the current acceleration with $w<-1$ might be just a start
of phantom inflation responsible for the observational universe in
upcoming cycle, and the universe recurs itself. It might be
interesting to see whether this scenario is altered by the
evolution of perturbation on large scale,
e.g.\cite{Piao0901},\cite{Zhang2010bb}.

It is generally difficult to foresee the future of our universe,
since there are lots of the evolutions or the parametrizations of
$w$, which may be consistent with the current observations but
have different aftertime, e.g.\cite{MZ}. However, if the universe
is recurring, the observations for the primordial perturbation
might help us to see the `tomorrow' of our universe.

Here, the recurring time is determined by the potential of the
phantom field. However, if there is a large step in its potential,
which straightly joint the current low energy regime to the regime
of phantom inflation, this time will be abridged. This might leave
an observational effect in upcoming universe, e.g. a lower CMB
quadrupole, which will be showed in later work.

\textbf{Acknowledgments}

We thank Sergei D. Odintsov, Taotao Qiu for helpful discussions.
This work is supported in part by NSFC under Grant No:10775180,
11075205, in part by the Scientific Research Fund of
GUCAS(NO:055101BM03), in part by National Basic Research Program
of China, No:2010CB832804

\appendix

\section*{Appendix:A}

The little rip phase will result in the disintegration of bound
structures, e.g.\cite{Atv}. The acceleration of the universe
brings an inertial force on a mass $m$ as felt by a gravitational
source separated by a comoving distance $l$,
\begin{eqnarray}
F_{in}=ml\frac{\ddot{a}}{a}={ml\over 6}(2\rho(a)+\rho^\prime(a)
a),
\end{eqnarray}
where $M_P^2/8\pi=1$ and the energy density $\rho(a)$ is the
function of the scale factor. It is convenient to introduce
dimensionless parameter
\begin{eqnarray} {\bar F}_{in}= {2\rho(a)+\rho^\prime(a) a\over \rho_0},  \label{Fi}\end{eqnarray}
where $\rho_0$ is the dark energy density at the present time. The
disintegration time of bound structures can be calculated by
applying Eq.(\ref{Fi}). In Ref.\cite{Atv}, it is found that the
system of Sun and Earth disintegrates when ${\bar F}_{in}$ reaches
$\sim 10^{23}$. Thus at this time
\begin{eqnarray} \rho(a)\sim 10^{23} \rho_0\sim 10^{-100}\ll \rho_{inf},\end{eqnarray}
where $\rho_{inf}\sim 10^{-12}$ is the inflationary energy
density. Thus the time before the phantom inflation is far larger
than the remaining time for the dissolution of bound structure.

\section*{Appendix:B}


We consider the potential as
\begin{eqnarray}\label{potential}
V_M(\phi)=\frac{1}{2}M^2\phi^2\left(1+\beta\tanh(\frac{\phi-\phi_{Step}}{\Delta})\right).
\end{eqnarray}
This potential has a step at $\phi_{Step}$. The parameters $\beta$
and $\Delta$ determine the height and width of this step. Though
there is an abrupt change of the evolution of the phantom field
due to the exist of step, the phantom field will climb up
continuously through the step while the phantom inflation will not
be ceased.



\begin{figure}[htbp]
\includegraphics[scale=0.6,width=8.5cm]{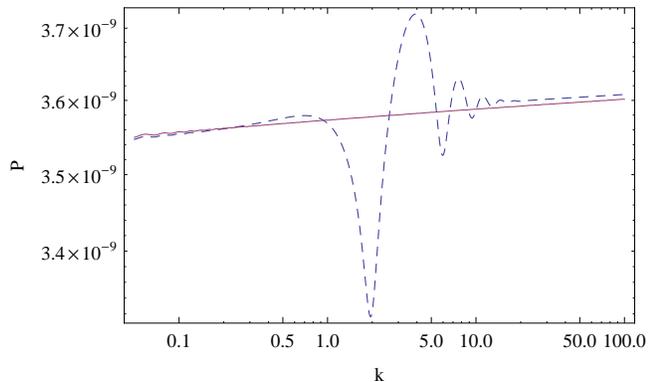}
\caption{The power spectrum of curvature perturbation for the
phantom inflation with the potential (\ref{potential}). The solid
and dashed lines are those without and with the step,
respectively. } \label{fig:m}
\end{figure}


%
\begin{figure}[htbp]
\includegraphics[scale=0.6,width=8.5cm]{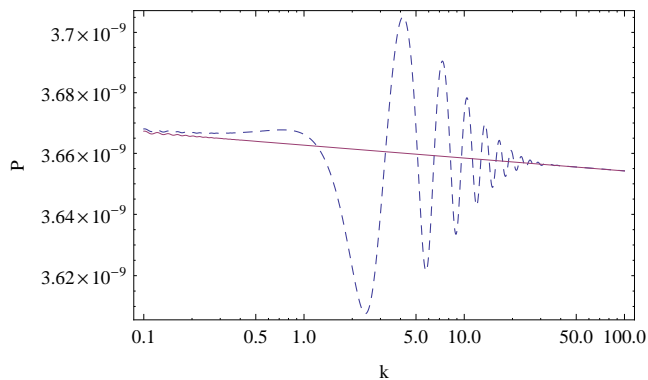}
\caption{The power spectrum of curvature perturbation for the
phantom inflation with the potential $V_0+V_M$. The solid and
dashed lines are those without and with the step, respectively.}
\label{fig:v0}
\end{figure}

The power spectrum for the phantom inflation with this potential
is plotted in Fig.\ref{fig:m}. The spectrum is scale invariant.
The step in the potential will result in a burst of oscillations,
and the magnitude and extent of oscillation are dependent on the
height and width of the step. These oscillations might provide a
better fit e.g.\cite{J3S},\cite{HAJ}.


The normal inflation model with the same potential has a red
spectrum $n_{\mathcal {R}}<1$, while the phantom inflation has a
blue spectrum $n_{\mathcal {R}}>1$, since \be \epsilon_{Pha}\simeq
\delta_{Pha}<0.\ee However, we can have the models of the phantom
inflation with $n_{\mathcal {R}}<1$. We consider an alternative
potential as $V_0+V_M$,
%
%
in which $V_M$ is Eq.(\ref{potential}) and $V_0$ is constant
dominating the potential. Here, \be |\epsilon_{Pha}|\ll
|\delta_{Pha}|, \ee and both are negative. Fig \ref{fig:v0} shows
that the spectrum is scale invariant, but has a slightly red tilt.

\end{document}